\documentclass{elsart}
\usepackage{graphics,natbib,graphicx,psfig,amssymb,ccaption} 
\usepackage{lscape,amsmath} 
\def\astrobj#1{#1}
\journal{New Astronomy}
\input psfig.sty
\begin{document}
\begin{frontmatter}
\title{Study of Dust and Ionized gas in Early-type Galaxies}
\author[Raipur]{Samridhi Kulkarni}
\ead{samridhikulkarni2005{@}gmail.com}
\author[Bangalore]{D. K. Sahu},
\author[Raipur]{Laxmikant Chaware},
\author[Raipur]{N. K. Chakradhari},
\author[Raipur]{S. K. Pandey}
\address[Raipur]{School of Studies in Physics \& Astrophysics,
Pt. Ravishankar Shukla University, Raipur (C. G.), 492010, India}
\address[Bangalore]{Indian Institute of Astrophysics,
 Koramangala, Bangalore 560034, India}

\begin{abstract}
We present results of optical broad-band and narrow-band H$\alpha$ observations of 
a sample of forty  nearby early-type galaxies. 
The majority of sample galaxies are known to have dust in various forms viz. dust lanes, nuclear dust and 
patchy/filamentary dust. A detailed study of dust was  performed for 12 galaxies 
with prominent dust features. The extinction curves for these galaxies run parallel to the 
Galactic extinction curve, implying that the properties of dust in these galaxies   are similar 
to those of the Milky-Way.  The ratio of total to selective extinction  ($R_{V}$) 
varies between 2.1 to 3.8, with an average of 2.9$\pm$0.2, fairly close to its 
canonical value of 3.1 for our Galaxy. The average relative grain size $\frac{<a>}{a_{Gal}}$ of  
dust particles in these 
galaxies turns out to be  1.01$\pm$0.2, while  dust mass  estimated using optical extinction 
 lies in the range $\sim$ $10^{2}$ to $10^{4}$ M$_{\odot}$.
The H$\alpha$ emission  was detected in 23 out of 29 galaxies imaged through narrow-band
filters with the H$\alpha$ luminosities in 
the range 10$^{38}$ - 10$^{41}$ erg sec$^{-1}$. The mass of the ionized gas is
in the range $\sim$  10$^{3}$ - 10$^{5}$ M$_{\odot}$. The morphology and extent  of  
ionized gas  is found similar to those of dust, indicating 
 possible coexistence  of dust and ionized gas in these galaxies. 
The absence of any apparent correlation between  blue luminosity and 
normalized IRAS dust mass is suggestive of merger related origin of dust and gas in these galaxies.
\end{abstract}
\end{frontmatter}

\section{Introduction}
Interstellar matter (ISM), being an important factor governing the formation and 
subsequent evolution of galaxies, has drawn considerable attention since their detection
and is still being continued to gain better understanding in various galactic and extragalactic
 environments. The availability of space observatories/satellites as well as ground-based 
telescopes with state of the art detectors  has made exploration of different 
phases of ISM  possible.

From the first detection of dust lanes in early-type galaxies 
(\citealt{Ber78}), the possibility of using its orientation  to extract 
additional information regarding  intrinsic shape of galaxies 
was explored (\citealt{Gun79, vanA82, Habe85, Habe88}). 
 Importance of dust in understanding  three dimensional structure 
of these galaxies led to search  for  dust in several of them 
 (\citealt{Haw81, Ebn85, Ver88, Sad85}).
Subsequently,  deep optical imaging surveys of large sample of early-type galaxies 
 detected dust in a significant fraction of them  
(\citealt{Goud94b, Fer99}).   Dust has also been
 detected in the innermost regions of early-type  galaxies in deep {\it Hubble Space Telescope} 
(HST) images, which remained  
unresolved with the ground-based observations (\citealt{Jaf94, van95,
deKoff00, Tran01}). Further, 
the detection of significant far-infrared (FIR) emission from early-type galaxies, 
using {\it Infra Red Astronomical Satellite} (IRAS),  
{\it Infrared Space Observatory} (ISO) and {\it Spitzer Space Telescope}
has added a new dimension in the study of dust and its possible role in 
underlying dynamics of galaxies (\citealt{Temi04, Xil04}).

  Optical broad-band imaging of galaxies with prominent dust features allows one to investigate physical 
properties of dust such as  particle size, extinction, reddening,  total dust content of galaxies 
and the processes that govern their evolution in different environments. 
Investigation of  dust extinction in different bands {\it i.e.} extinction curve
 has been traditionally  a basic tool for studying  dust properties.
A refined form of this technique has been applied to study the  properties of dust in a  
number of dusty early-type galaxies (\citealt{Goud94a, Sahu98, Pat07, Fin08, Fin10d}). 
 These studies showed 
that the extinction curves of dust in early-type galaxies run almost parallel 
to that of the Milky-Way, with average relative grain size not very much different 
from that in our Galaxy. 

Optical spectroscopic observations of early-type galaxies show that $\sim$55 - 60\% 
of them have faint emission line,  indicating the presence of  ionized gas, 
with mass $\sim$ 10$^{3}$ -10$^{4}$ M$\odot$  (Caldwell 1984, Phillips et al. 1986). 
Further, the H$\alpha$ imaging survey of large sample of early-type galaxies confirmed the 
presence of  ionized gas with various morphologies; such as flattened disc, ring or
filamentary structures  (\citealt{Kim89, Shi91, Trin91, Bus93, Goud94a, Sing95, Macch96, Mart04}). 
The possible sources of gas excitation   have been explored and 
the post-asymptotic giant branch (pAGB) stars  were identified as the main contributor of 
ionizing radiation. However,  for at least  10\% of  early-type galaxies,  the ionized
emission is powered by  recently formed stellar subcomponent (\citealt{Sarzi10}).   

Other forms of ISM,  {\it i.e.} hot and cold  gas have also been found in early-type galaxies.  
Hot gas halos have been detected  around early-type galaxies
with  X-ray observatories  (\citealt{For85, Can87, Fab89, Osul01, Sar01, Kim03, Kim10}). 
Cold molecular gas is detected through
CO emission in $\sim$ 22\% of all early-type galaxies (\citealt{You11} and references therein).  

In a large fraction of dusty early-type galaxies, the morphology and extent of   ionized gas 
match with that of dust (\citealt{Goud94b, Fer99, Pat07}) and in some cases
with the X-ray emitting region too (\citealt{Goud98}). This points towards a possible physical 
connection between hot, warm and cold phases of ISM.
 The origin and fate of ISM in early-type galaxies is important as it holds clue to the formation
and subsequent evolution of early-type galaxies   (\citealt{Goud95}). The recent studies have shown 
that  neither internal nor external origin can explain all the observed properties of ISM in these 
galaxies.  Hence, a  good balance between internal and external origin is suggested as the
source of ISM in early-type galaxies.

In the present paper,  we discuss the properties of dust and ionized gas in a sample of forty  low 
redshift early-type galaxies.  The paper is organized as follows: Observation and data 
reduction  in Section 2. Section 3 gives the methodology used for analyzing 
dust and ionized gas, the results have been discussed in Section 4.  We summarize our results in Section 5.
\begin{table}
\begin{scriptsize}
\begin{center}
\caption{Global parameters of sample galaxies}
\begin{tabular}{|lllllcc|}
\hline
 Object  & RA       & DEC       & Morph.    &$B^{0}_{T}$& $V_{Helio}$& Size \\
         &(J2000)   &(J2000)    &(RC3)      &           &  (km/s)  & (arcmin)\\
\hline
NGC\,0383& 01:04:39 & \,36:09:07 & SA0  & 13.38 & 5098 &  1.6x1.4 \\
NGC\,0708& 01:52:46 & \,36:09:07 & E    & 13.70 & 4855 &  3.0x2.5 \\
NGC\,0720& 01:53:00 & -13:44:19  & E5   & 11.16 & 1745 &  4.7x2.4 \\
NGC\,1052& 02:41:04 & -08:15:21  & E4   & 12.10 & 1510 &  3.2x2.1 \\
NGC\,1167& 03:01:42 & \,35:12:21 & SA0  & 13.38 & 4945 &  2.8x2.3 \\
NGC\,1199& 03:01:18 & -15:48:29  & E3   & 12.37 & 2570 &  2.4x1.9 \\
NGC\,1395& 03:38:29 & -23:01:40  & E2   & 10.97 & 1717 &  5.9x4.5 \\
UGC\,2783& 03:34:18 & \,39:21:25 & S0   & 12.99 & 6173 &  1.3x1.2 \\
NGC\,1407& 03:40:11 & -18:34:49  & E0   & 10.70 & 1779 &  4.6x4.3 \\
NGC\,2534& 08:12:54 & \,55:40:19 & E1   & 13.70 & 3447 &  1.4x1.2 \\
NGC\,2644& 08:41:31 & \,04:58:49 & S    & 13.31 & 1939 &  2.1x0.8 \\
NGC\,2768& 09:11:37 & \,60:02:14 & S0   & 10.84 & 1373 &  8.1x4.0 \\
NGC\,2851& 09:20:30 & -16:29:43  & E    & 15    & 5195 &  1.2x0.5 \\
NGC\,2855& 09:21:27 & -11:54:34  & SA   & 12.63 & 1897 &  2.5x2.2 \\
NGC\,3065& 10:01:55 & \,72:10:13 & SA   & 13.5  & 2000 &  1.7x1.7 \\
NGC\,3115& 10:05:14 & -07:43:07  & S0   & 09.87 & 663  &  7.2x2.0 \\
NGC\,3377& 10:47:42 & \,13:59:08 & E5   & 11.24 & 665  &  5.2x3.0 \\
M\,105   & 10:47:49 & \,12:34:54 & E1   & 10.24 & 911  &  5.4x4.0 \\
NGC\,3489& 11:00:18 & \,13:54:04 & S    & 11.12 & 677  &  3.5x2.0 \\
NGC\,3607& 11:16:54 & \,18:03:07 & SA   & 10.82 & 960  &  4.9x2.5 \\
NGC\,3801& 11:40:16 & \,17:43:41 & S0   & 12.96 & 3317 &  3.5x2.1 \\
NGC\,4125& 12:08:06 & \,65:10:27 & E6   & 10.65 & 1356 &  5.8x3.2 \\
NGC\,4233& 12:17:07 & \,07:37:28 & S0   & 12.8  & 2371 &  2.3x0.9 \\
NGC\,4278& 12:20:06 & \,29:16:51 & E1   & 11.20 & 649  &  4.1x3.8 \\
NGC\,4365& 12:24:28 & \,07:19:03 & E3   & 10.52 & 1243 &  6.9x5.0 \\
NGC\,4494& 12:31:24 & \,25:46:30 & E1   & 10.71 & 1344 &  4.8x3.0 \\
NGC\,4552& 12:35:39 & \,12:33:23 & E    & 10.73 & 340  &  5.1x4.7 \\
NGC\,4648& 12:41:44 & \,74:25:15 & E3   & 12.96 & 1414 &  2.1x1.6 \\
NGC\,4649& 12:43:39 & \,11:33:09 & E2   & 09.81 & 1117 &  7.4x6.0 \\
NGC\,4697& 12:48:35 &  -05:48:02 & E6   & 10.10 & 1241 &  7.2x4.7 \\
NGC\,4874& 12:59:35 & \,27:57:34 & cD   & 12.63 & 7224 &  1.9x1.9 \\
NGC\,5322& 13:49:15 & \,60:11:26 & E3   & 11.14 & 1754 &  5.9x3.9 \\
NGC\,5525& 14:15:39 & \,14:16:57 & S0   & 13.6  & 5553 &  1.4x0.9 \\
NGC\,5812& 15:00:55 &  -07:27:26 & E0   & 12.19 & 1970 &  2.1x1.9 \\
NGC\,5846& 15:06:29 & \,01:36:20 & E0   & 11.05 & 1714 &  4.1x3.8 \\
NGC\,5866& 15:06:29 & \,55:45:48 & S0   & 10.74 & 672  &  4.7x1.9 \\
NGC\,6166& 16:28:38 & \,39:33:06 & E2   & 12.78 & 9100 &  1.9x1.4 \\
NGC\,7052& 21:18:33 & \,26:26:48 & E    & 13.40 & 4672 &  2.5x1.4 \\
NGC\,7454& 23:01:07 & \,16:22:58 & E4   & 12.78 & 2022 &  2.2x1.6 \\
IC\,2476 & 09:27:52 & \,29:59:09 & S0   & 13.85 & 8007 &  1.5x1.4\\
\hline
\multicolumn{7}{p{12cm}}{ Cols.(2) and (3) list galaxy co-ordinates,
 Col.(4) lists morphological classification of the galaxies,blue luminosity of 
the program galaxies are listed in Col.(5), while Col.(6) lists heliocentric velocity,
Col.(7) optical size of the galaxies, all taken from RC3(de Vaucoulers et al. 1991)}
\end{tabular}
\label{basic}
\end{center}
\end{scriptsize}
\end{table}

\section{Data acquisition and reduction}
\subsection{Observation}
The  aim  of this work is to study the properties of dust and its relationship with 
 ionized gas,  in a large sample of early-type galaxies containing  dust features. 
Our sample consists of 40 nearby (z $<$ 0.02) early-type galaxies (E/S0)  harboring 
some form of dust. Our program galaxies were selected from   \cite{Ebn85}, \cite{BB98}, \cite{Gon00}
and \cite{Tran01}. Apart from the redshift 
constraint and possible presence of dust in them, no strict criterion was applied
for selecting the sample. The program galaxies were chosen depending on the
availability of the observing time.  The basic details of the sample galaxies taken 
from the RC3 catalog \cite{dVau92}  are listed in Table 1.\\
The observations were carried out  during  2004 to 2010,
using 2-m Himalayan Chandra Telescope (HCT) of Indian Astronomical Observatory (IAO), 
Hanle  and 2-m telescope at IUCAA Girawali Observatory (IGO), 
Pune. Observations were made using HFOSC and IFOSC mounted on HCT and IGO, 
 respectively. The details of the instrument used for observations in various runs are 
listed in Table 2. 
 The gain and readout noise of the 
CCDs used   are 1.22 $e^{-}/ADU$ and 4.87 $e^{-}$ at HCT and 1.4 $e^{-}/ADU$ and 10 $e^{-}$ 
at IGO. 
The seeing during our observations was between  1.0  and 2.5 arcsec. 
Broad-band imaging  in   Bessell's $B, V, R$ and $I$ filters for all 
the program galaxies  were carried out with the HCT. 
 The effective exposure time was chosen in such a way  to get images with good and 
approximately similar signal-to-noise ratio  in each band. 
Exposure time was split into several short exposures to avoid saturation of
 foreground stars and proper removal of the cosmic-ray hits.
Several bias frames and twilight sky frames were also taken for pre-processing 
of the data.
\begin{table}\tiny{
\caption{Details of observing runs}
\begin{tabular}{|lllllll|}
\hline
Run & Observatory & Instrument & Format & Scale & FOV & Filters used\\
    &             &           &of CCD &(''/pixel) &&\\
\hline
April 2004 & IAO  & HFOSC     &  2k $\times$ 2k & 0''.29 & 10' $\times$ 10' & $B, V, R, I$, H$\alpha$\\
Oct. 2004  & IAO  & HFOSC     &  2k $\times$ 2k & 0''.29 & 10' $\times$ 10' & $B, V, R, I$, H$\alpha$\\
Dec. 2004  & IAO  & HFOSC     &  2k $\times$ 2k & 0''.29 & 10' $\times$ 10' & $B, V, R, I$, H$\alpha$\\
Jan. 2005  & IAO  & HFOSC     &  2k $\times$ 2k & 0''.29 & 10' $\times$ 10' & $B, V, R, I$, H$\alpha$\\
March 2005 & IAO  & HFOSC     &  2k $\times$ 2k & 0''.29 & 10' $\times$ 10' & $B, V, R, I$, H$\alpha$\\
May 2005   & IAO  & HFOSC     &  2k $\times$ 2k & 0''.29 & 10' $\times$ 10' & $B, V, R, I$\\
Nov. 2005  & IAO  & HFOSC     &  2k $\times$ 2k & 0''.29 & 10' $\times$ 10' & $B, V, R, I$\\
Dec. 2005  & IAO  & HFOSC     &  2k $\times$ 2k & 0''.29 & 10' $\times$ 10' & $B, V, R, I$\\
Mar. 2008  & IAO  & HFOSC     &  2k $\times$ 2k & 0''.29 & 10' $\times$ 10' & $B, V, R, I$\\
           & IGO  & IFOSC     &  2k $\times$ 2k & 0''.3  & 10' $\times$ 10' & 6563, 6603, 6683\\
April 2008 & IAO  & HFOSC     &  2k $\times$ 2k & 0''.29 & 10' $\times$ 10' & $B, V, R, I$\\
           & IGO  & IFOSC     &  2k $\times$ 2k & 0''.3  & 10' $\times$ 10' & 6563, 6603, 6683\\
Aug. 2008  & IAO  & HFOSC     &  2k $\times$ 2k & 0''.29 & 10' $\times$ 10' & $B, V, R, I$\\
Feb. 2009  & IAO  & HFOSC     &  2k $\times$ 2k & 0''.29 & 10' $\times$ 10' & $B, V, R, I$\\
April 2009 & IGO  & IFOSC     &  2k $\times$ 2k & 0''.3  & 10' $\times$ 10' & 6563, 6603, 6683\\
Oct. 2009  & IGO  & IFOSC     &  2k $\times$ 2k & 0''.3  & 10' $\times$ 10' & 6563, 6603, 6683\\
Dec. 2009  & IAO  & HFOSC     &  2k $\times$ 2k & 0''.29 & 10' $\times$ 10' & $B, V, R, I$\\
Jan. 2010  & IAO  & HFOSC     &  2k $\times$ 2k & 0''.29 & 10' $\times$ 10' & $B, V, R, I$\\
           & IGO  & IFOSC     &  2k $\times$ 2k & 0''.3  & 10' $\times$ 10' & 6563, 6603, 6683\\
\hline
\end{tabular}
\label{run}
}
\end{table}

\begin{table}\begin{center}\tiny{
\caption{Characteristics of H$\alpha$ filter}
\begin{tabular}{|cccc|}
\hline
Filter Name & $\lambda_{cent.}$ & Band-pass & Peak tran.\\
&&$\dot{A}$&$\%$\\
\hline
HCT-6563 & 6563 & 100 & 86$\%$\\
IGO-6563 & 6563 & 80  & 90$\%$\\
IGO-6603 & 6603 & 80  & 89$\%$\\
IGO-6683 & 6683 & 80  & 88$\%$\\
\hline
\end{tabular}
\label{h_filter}
}\end{center}
\end{table}
 Twenty nine out of 40 sample galaxies  were also observed
 in narrowband H$\alpha$  filters.  The H$\alpha$ filter available with the HFOSC is centered
at 6563 \AA \ with a band-pass of 100 \AA, so relatively nearby galaxies were observed
in  H$\alpha$ with the HCT.  The $R$ band image was used for removing  the overlapping
continuum as suggested by \cite{Wal90}. The IFOSC is equipped with redshifted H$\alpha$ filters.  
The H$\alpha$ filter was selected according to the redshift of the galaxy and images
through the adjacent narrow-band  filter was taken for continuum subtraction.
The details of the narrow-band  filters used are listed in Table 3. 
Spectrophotometric standards, Hiltner 600, Feige 34
and GD140 from the list of \cite{Oke74}, were also observed for absolute flux calibration. 
Details of the observations are listed in Table 4. 
\subsection{Data Reduction}
The  Image Reduction and Analysis Facility
(IRAF{\footnote {IRAF is distributed by the National Optical Astronomy Observatories, 
which are operated by the Association of Universities for Research in Astronomy, Inc., 
under cooperative agreement with the National Science Foundation.}}) was used
for data reduction.  The pre-processing  such as bias subtraction, flat-fielding  
  was carried out using various tasks available within IRAF in a standard manner. 
Multiple frames taken in each filter were
geometrically aligned to an accuracy better than one tenth of a pixel using
 {\it`geomap'} and {\it`geotran'}, the aligned frames were  combined to 
generate a final image with improved  S/N ratio. This also enabled easy removal of cosmic ray events. 
Sky value in galaxy frame was measured within a box of size close to  full width at half maximum 
of stellar profile, at several locations away from the galaxy.  An average of these 
values was taken as the sky level. The sky value was also
estimated by fitting a powerlaw to the outer parts of the galaxy as described in
\citet{Goud94b}, results of both the methods were found to agree  well.
   Cleaned, sky subtracted images of the
program galaxies  were  used for further analysis.
\begin{table}\begin{center}\tiny{\bf \it
\caption{Log of observation }
\begin{tabular}{|l|lllllll|}
\hline
Galaxy &\multicolumn{6}{c}{Total Exposure time(sec)}& ${H\alpha}$ data\\
Name&B&V&R&I&{H$\alpha$}&{Cont.}&HCT/IGO\\
\hline

NGC\,0383 & 1500 & 720  & 540 & 720 &  NA &  &  - \\
NGC\,0708 & 3000 & 1200 & 1200& 1200& 5400 & 3600& IGO  \\
NGC\,0720 & 2160 & 1080 & 900 & 900 & 3600 & 900 & HCT  \\
NGC\,1052 & 2100 & 1260 & 360 & 480 & 4800 & 4800& IGO  \\
NGC\,1167 & 2700 & 3150 & 2220& 1200& 3600 & 900 & IGO  \\
NGC\,1199 & 1800 & 900  & 900 & 720 & 5700 & 900 & HCT  \\
NGC\,1395 & 2400 & 840  & 540 & 450 & 2700 & 540 & HCT  \\
UGC\,2783 & 3000 & 900  & 960 & -   & 1800 & 960 & IGO  \\
NGC\,1407 & 2280 & 1200 & 1080& 720 & 4500 & 1080& HCT  \\
NGC\,2534 & 1800 & 900  & 600 & 600 & 6300 & 3600& IGO  \\
NGC\,2644 & 2400 & 1680 & 1020& 900 & 7200 & 4800& IGO  \\
NGC\,2768 & 2100 & 720  & 720 & 720 & 4200 & 720 & HCT  \\
NGC\,2851 & 2700 & -    & 900 & 1200& NA   &     & -  \\
NGC\,2855 & 2400 & 2520 & 840 & 720 & 5400 &3600 & IGO  \\ 
NGC\,3065 & 3480 & 1920 & 1500& 1500& NA   &     & -  \\
NGC\,3115 & 1380 & 480  & 180 & 60  & 900  & 180 & HCT  \\ 
NGC\,3377 & 1680 & 900  & 600 & 510 & 2700 & 600 & HCT  \\
M\,105    & 900  & 480  & 180 & -   & 3600 & 180 & HCT  \\
NGC\,3489 & 600  & 1200 & 900 & 900 & 5400 & 2700& IGO  \\
NGC\,3607 & 1800 & 900  & 600 & -   & 6600 & 600 & HCT  \\
NGC\,3801 & 2700 & 1800 & 800 & 1200& 7800 & 800 & HCT  \\
NGC\,4125 & 2580 & 780  & 1200& 2430& 1800 &1200 & HCT  \\      
NGC\,4233 & 4200 & 2700 & 2100& 2040& NA   &     & -  \\
NGC\,4278 & 2100 & 1000 & 1200& 900 & NA   &     & -  \\
NGC\,4365 & 2400 & 1200 & 1980& 2400& 3000 & 1980& HCT  \\
NGC\,4494 & 2400 & 1080 & 840 & 540 & NA   &     & -  \\
NGC\,4552 & 900  & 1350 & -   & -   & NA   &     & -  \\
NGC\,4648 & 3300 & 2640 & 1980& 900 & NA   &     & -  \\
NGC\,4649 & 2280 & 1290 & 1440& -   & 7500 & 1440& HCT  \\      
NGC\,4697 & 1200 & 780  & 600 & 480 & 1800 & 600 & HCT  \\
NGC\,4874 & 2400 & 1200 & 600 & 540 & NA   &     & -  \\
NGC\,5322 & 1200 & 600  & -   & -   & NA   &     & -  \\
NGC\,5525 & 3600 & 1500 & 1800& 2700& 3600 & 1200& IGO  \\
NGC\,5812 & -    & 3000 & 1500& -   & NA   &     & IGO  \\
NGC\,5846 & 1800 & 600  & 1470& 1200& 2400 & 1200& IGO  \\
NGC\,5866 & 2400 & 1040 & 1200& 720 & 3000 & 1200& HCT  \\
NGC\,6166 & 1800 & 900  & 540 & 540 & 1800 & 1620& IGO  \\
NGC\,7052 & 2400 & 1320 & 1260& 1200& 5400 & 3600& IGO  \\
NGC\,7454 & 1440 & 570  & 105 & -   & 4500 & 900 & IGO  \\       
IC\,2476  & 900  & 600  & 600 & -   & NA   &     & -  \\
\hline
\end{tabular} 
\label{observation}
}\end{center}
\end{table}

\section{Analysis}
\subsection{Broad-band images}
 Most of our sample galaxies are known to harbor dust in them, however, information 
regarding presence of dust in some galaxies  selected from the X-ray sample of \cite{BB98}, 
and radio sample of  \cite{Gon00}  is not 
available. So,  first an attempt was made to look for possible presence of dust in these 
galaxies. 
Various image processing techniques such as colour index map, quotient image, 
unsharp masking etc. were used, 
to identify dust feature, its morphology and extent of galaxy affected by it. 
\subsubsection{Colour Maps} 
The $(B-V)$ and $(B-R)$ colour index map of the sample galaxies were created using 
processed  and point spread function (psf) matched  
broad-band images.  Any difference in  psf in the broad-band images 
may lead to detection of spurious features in the colour maps. Hence, the image with better 
seeing was convolved with a Gaussian kernel to match with the psf of images in other band. 

 While generating colour maps \cite{Goud94c}  have avoided
using $R$ filter, as H$_\alpha$ and [NII] emission, if present in the galaxy, lies 
within the  $R$ filter. It is shown by \cite{Fer99} that due to the wide band-pass of  
$R$ filter ($\sim$1500 \AA) and relatively small equivalent width of the H$_\alpha$ + [NII] 
emission in early-type galaxies, the contamination introduced by H$_\alpha$ + [NII]  
emission  in the  $(B-R)$ colour will be less than 1\% or 0.01 mag,
and hence may be neglected. As our $I$ band images suffer 
 from  fringing  (due to use of thin CCDs) and we donot have $I$ band image 
for all the sample galaxies,  we opted for  $(B-V)$ and $(B-R)$ colour images.
The  colour index map of sample galaxies are displayed  in Figure 1. 
 Dust is detected in 35 out of 40 sample galaxies.   We report the presence of dust 
in  two galaxies namely, \astrobj{NGC 4649} and \astrobj{NGC 4874} for the first time. 
Further, we confirm the  presence of dust in 33 galaxies.
In our sample,  12 galaxies show dust in the form of   prominent ring, lane or filaments,   
whereas in other cases nuclear dust  patch is detected.  
\begin{figure}
\begin{center}
\includegraphics[totalheight=8in,width=5in]{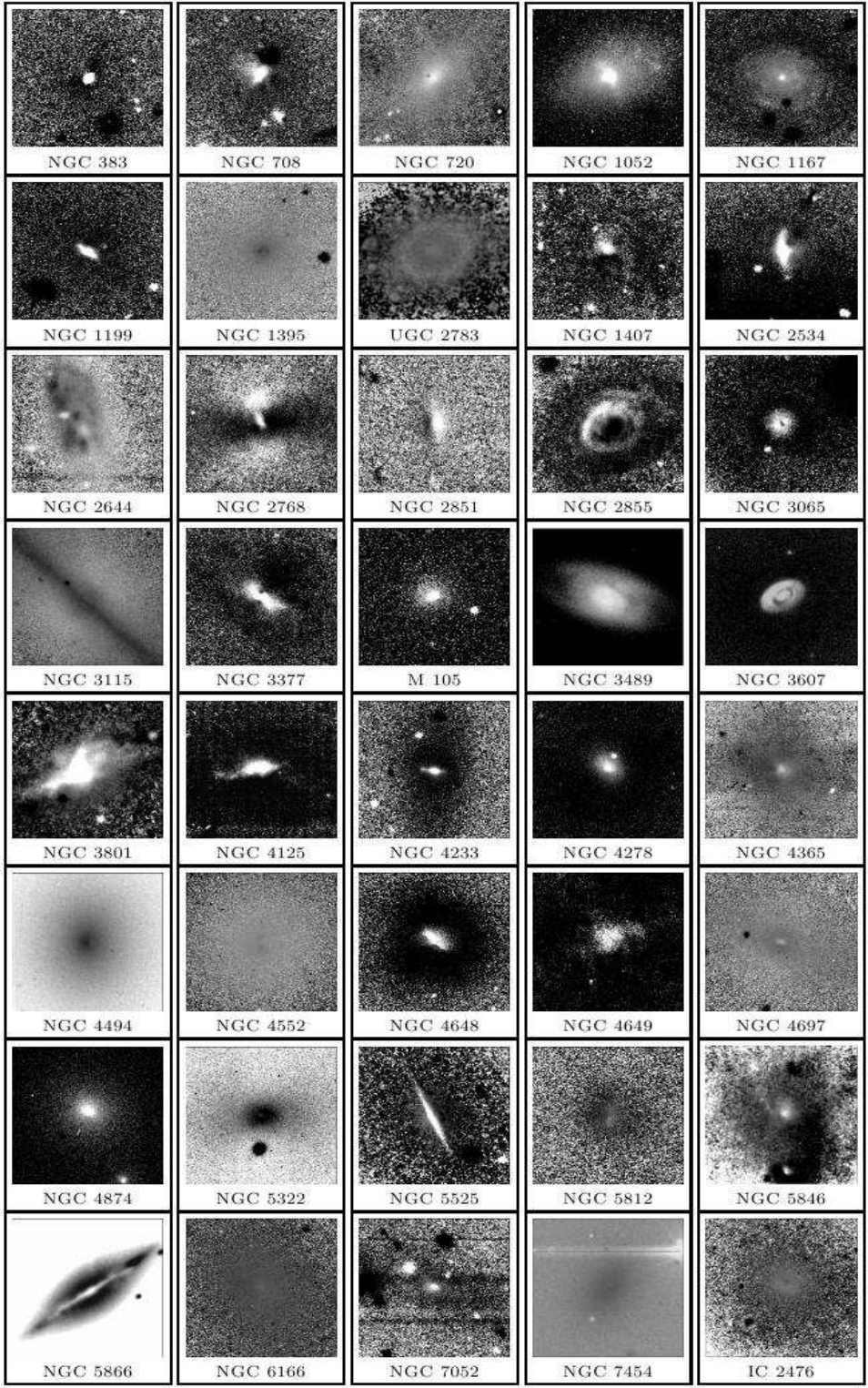}
\caption{$(B-R)$ Colour maps for sample Galaxies. For \astrobj{NGC 720}, \astrobj{UGC 2783}, \astrobj{NGC 3065},
\astrobj{NGC 4278} and \astrobj{NGC 5812} feature was not seen clearly in $(B-R)$ we have shown $(V-R)$. 
Brighter shades represent dust occupied region. 
North is up and east is to the left.}
\label{color}
\end{center}
\end{figure}

\subsubsection{Extinction Maps} 
From the isophotal contours and surface brightness maps, it appears that  in absence of 
any embedded feature,  stellar light distribution of early-type galaxies  is  smooth.
The isophotal contours of these galaxies can be well sampled by fitting ellipse on them.
 Presence of  non-elliptical features such as dust lanes, dust patches,  disk 
in these galaxies can distort the isophotes and they are no more perfect ellipses.

The amount of extinction caused
by the dust present in galaxy is determined by comparing the light
distribution of the galaxy with that expected in absence of
dust. For this, ellipse fitting was carried out on the processed galaxy images 
using {\it `ellipse'} task available in {\it `stsdas'} package in IRAF,  in an iterative way
 as  mentioned by \cite{Pat07}.  The ellipse fitting was first carried out for 
$R$ and $I$ band images which are least affected by  the dust extinction. The center 
coordinate of the ellipses was determined 
by averaging those of the best fitted ellipses in $R$ and $I$ band images. This  
was kept fixed for fitting ellipses to the images in other bands.  
During the next iteration of ellipse fitting,  the dust occupied
regions identified with the colour index maps, were masked off to avoid the 
effect of distortion introduced by  dust. 
As the extent of dust features are small  compared to the size of the galaxy,  masking it 
does not affect the fit. 
 The deviations of isophotes from perfect ellipses, if any,  are reflected in 
the higher order harmonics  $a3$, $a4$ and $b3$, $b4$.  These are the amplitudes of $\sin 3\theta$, 
$\sin 4\theta$ and  $\cos 3\theta$,  $\cos 4\theta$ coefficients of the isophotal 
deviation from perfect  ellipse. In the presence of features which distort 
the isophotes,  the intensity distribution of  early-type 
 galaxies can be modeled well by including higher order 
harmonics in the fit.
A smooth, 
dust free model image of the galaxy  was created by applying polynomial fit to 
 the isophotal parameters generated by the
ellipse fitting process  and including the higher order harmonics using {\it `bmodel'} task. 
Extinction maps in different bands  were generated using the 
smooth dust free model of the galaxies in the following way, 
\begin{equation}
A_{\lambda}=-2.5\times\log{\frac{I_{\lambda,obs}}{I_{\lambda,model}}}
\end{equation}
where A$_{\lambda}$ represents the amount of extinction in a given band ($B, V, R, I$),
and I$_{\lambda}$ stands for the ADU counts.  
Figure 2  
shows the extinction maps for galaxies with prominent dust feature.
The almost similar morphology of  dust features seen in the extinction and colour index 
 maps  supports  that the identified dust features are real and not the artifact 
introduced by the fitting procedure. 
These extinction maps are then used for making extinction curves and
calculating dust mass.
\begin{figure}
\begin{center}
\includegraphics[totalheight=3.4in,width=4in]{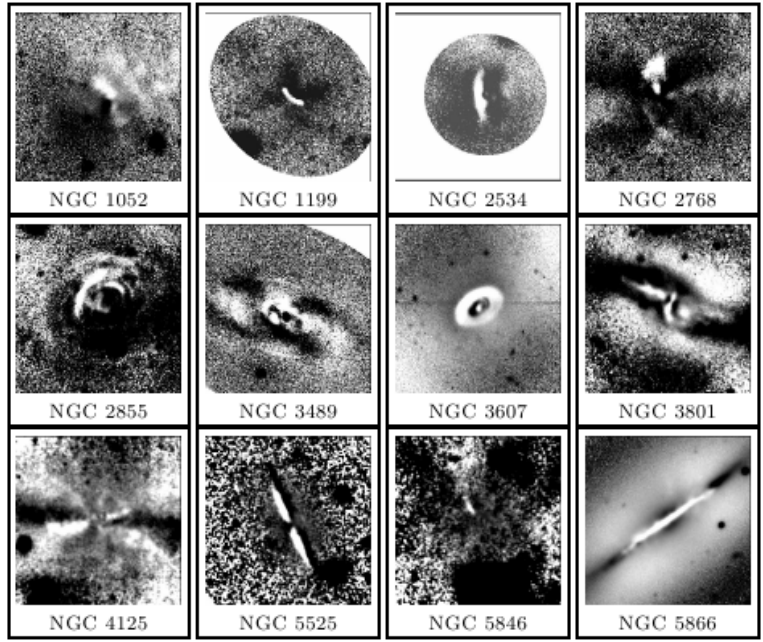}
\caption{ Extinction map in $B$ band for sample galaxies with prominent dust features.}
\label{extmap}
\end{center}
\end{figure}

\subsubsection{Extinction Curves}
The extinction and colour index maps reveal that 12 galaxies from our sample 
show  prominent dust features. An attempt is made to study  properties of dust 
viz. total extinction due to dust, extinction curve, 
relative particle size and dust mass for these galaxies. 
 The  total extinction $A_{\lambda}$ due to dust present is the only
  observed quantity, all other properties are directly related to it and hence 
proper care should be taken to determine $A_{\lambda}$. For determining $A_{\lambda}$, we adopted  the method
described by \cite{Goud94c} and \cite{Pat07}. The numerical value of total 
extinction $A_{\lambda}$ (${\lambda}$ = $B, V, R$ and $I$),  in each band was estimated 
by sliding a box of size comparable to that of seeing, over the dust affected part 
in the extinction map. 
The local value of selective extinction as a function of position  across 
the dust occupied region was derived  using the  values of 
extinction $A_{\lambda}$ measured  at different locations in individual band as 
$E(\lambda - V) = A_{\lambda} - A_{V}$. 
The scatter within each box was used to estimate the 
uncertainty associated with the extinction values.

A linear regression fit was made between various local values of total  extinction 
$A_{\lambda}$ (${\lambda}$ = $B, V, R$ and $I$).  The best fitting slope was assigned to be 
the average  slope of $A_{x}$ versus $A_{y}$  and the reciprocal slope of A$_{y}$ 
versus A$_{x}$ (where $x, y = B, V, R$,  $I$ and $x \neq y$). A similar fit was performed between 
the total extinction $A_{\lambda}$  and the  selective extinction
$E(B - V)$, the slope gives the ratio of total to selective extinction 
($R_{\lambda} = \frac{A_{\lambda}}{E(B - V)}$) for  a given  band (\citealt{Goud94c, Sahu98, Pat07}). 
The values of $R_{\lambda}$ thus derived for  galaxies with prominent dust features 
 are given in Table 5.  The 
$R_{\lambda}$ values for  the Milky Way taken from \cite{Rie85} are 
also listed for comparison.   

In order to study the wavelength dependence of extinction values, we plotted
$R_{\lambda}$ against $\lambda^{-1}$,  known as `extinction curve'  and compared it 
with that of the Milky Way in Figure 3. 
 The $R_{\lambda}$ varies linearly 
with inverse of  wavelength ($\lambda^{-1}$), 
consistent with the fact that extinction efficiency $Q_{ext}$ is proportional to 
${\lambda}^{-1}$ for small grain size \textit{x} $<$ 1, where $x = (\frac{2{\pi}a}{\lambda}$) 
and \textit{a}  is  grain radius. 
As seen in Figure 3 
 the  extinction curves derived for 
dusty  galaxies run parallel to that of the canonical curve of our Galaxy. This 
suggests that the  optical extinction properties of dust grains in extragalactic
environment are  similar to  those  of the Milky Way. 
\cite{Fin08}  explored various factors which can give rise to 
smaller or larger values of $R_{\lambda}$. They concluded that increasing (decreasing) the
number of upper end size grains with respect to the Galactic grain population produces 
larger (smaller) $R_{\lambda}$ values. The varying abundance ratio 
$\frac{A_{silicate}}{A_{graphite}}$ of the 
silicate and graphite grains  can also give rise to different values of $R_{\lambda}$.
Since the observed abundance of silicate and graphite in the ISM
is approximately equal, the variation in derived $R_{\lambda}$
 from that of the Milky Way, could only be explained due to
difference  in grain size. The relative
dust grain size $\frac{<a>}{a_{Gal}}$ is estimated by shifting the extinction
curve  as described by \cite{Hil83}, \cite{Goud94c} and  \cite{Fin08}. The relative
 dust grain size for the sample galaxies is given in Table 5.
\begin{figure}
\begin{center}
\includegraphics[totalheight=4in,width=3.8in]{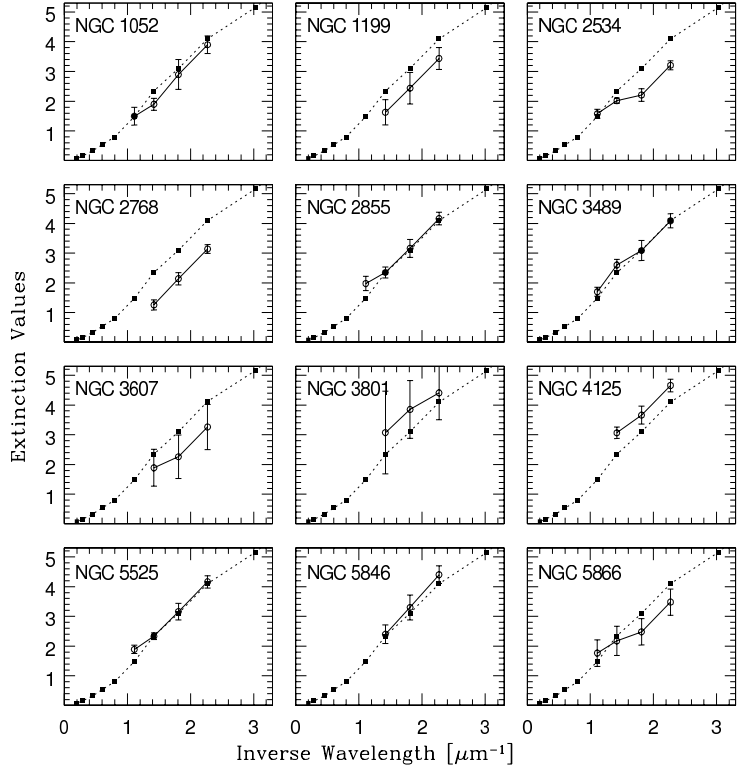}
\caption{Extinction curves for galaxies with prominent dust features (solid line).
Extinction curve for the Milky Way (dotted line) is also plotted for comparison}
\label{ext_cur}
\end{center}
\end{figure}

\begin{table}\begin{center}
\tiny{
\caption{$R_{\lambda}$ values, relative grain size and mass of dust}
\begin{tabular}{|lcccccc|}
\hline
Object  &$R_{B}$&$R_{V}$&$R_{R}$&$R_{I}$&$\frac{<a>}{a_{Gal}}$& log M$_{d,Opt}$\\
NGC&&&&&&M$_{\odot}$\\
\hline
1052&3.9$\pm$0.3&2.9$\pm$0.5&1.9$\pm$0.2&1.5$\pm$0.3&0.94&1.8$\pm$0.02\\   
1199&3.5$\pm$0.3&2.5$\pm$0.4&2.2$\pm$0.2&1.7$\pm$0.2&0.93&2.0$\pm$0.02\\      
2534&3.1$\pm$0.1&2.1$\pm$0.2&2.0$\pm$0.0&1.1$\pm$0.1&0.83&3.3$\pm$0.03\\      
2768&3.1$\pm$0.1&2.1$\pm$0.2&1.3$\pm$0.1&-&0.75           &2.0$\pm$0.01\\      
2855&4.2$\pm$0.2&3.2$\pm$0.3&2.4$\pm$0.1&1.9$\pm$0.2&0.95&3.1$\pm$0.02\\      
3489&4.1$\pm$0.2&3.1$\pm$0.3&2.6$\pm$0.1&1.7$\pm$0.1&1.04&2.4$\pm$0.05\\      
3607&3.3$\pm$0.7&2.3$\pm$0.7&1.9$\pm$0.6&-&0.87           &2.1$\pm$0.04\\      
3801&4.4$\pm$0.9&3.8$\pm$0.9&3.1$\pm$1.3&-&1.47           &4.0$\pm$0.09\\      
4125&4.6$\pm$0.2&3.6$\pm$0.2&3.1$\pm$0.1&-&1.36           &2.1$\pm$0.02\\      
5525&4.2$\pm$0.2&3.2$\pm$0.2&2.3$\pm$0.1&1.9$\pm$0.1&1.01&4.0$\pm$0.09\\      
5846&4.4$\pm$0.2&3.4$\pm$0.4&2.4$\pm$0.3&-&1.1            &2.8$\pm$0.02\\      
5866&3.4$\pm$0.4&2.5$\pm$0.4&2.2$\pm$0.4&1.8$\pm$0.4&0.98&3.0$\pm$0.09\\
&&&&&&\\
Galaxy&4.1&3.10&2.27&1.86&1.00&\\
\hline

\end{tabular}
\label{optical_mass}
}
\end{center}
\end{table}

\subsubsection{Dust mass from total extinction values} 
To estimate dust mass in galaxies with prominent dust features, we made use of 
the two component model described by \cite{Goud94c}. Assuming that the
 dust grains are spherical with radius \textit{a}, the extinction cross-section at  wavelength 
$\lambda$ is given as  
\begin{equation}
C_{ext} = \int^{a+}_{a-} Q_{ext}(a,\lambda) \pi a^{2} n(a) da
\end{equation}
where $Q_{ext}(a,\lambda)$ is the extinction efficiency at wavelength $\lambda$ and 
\textit{n(a)}  is the grain size 
distribution function defined as $n(a) = n_{0} a^{-3.5}$ for $a_{-} \leqslant a \leqslant a_{+}$,  
 $a_{-} = 0.005 \ \mu$m  and $a_{+}$=  0.22 $\mu$m are the 
lower and upper cut-offs of the grain size distribution, respectively
(\citealt{Mat77}, \citealt{Dra84}). Under the 
assumption that the grain size distribution function \textit{n(a)} is the same  
over the dusty region and $l_{d}$ is the 
dust column length along the line of sight, the  total extinction $A_{\lambda}$ is calculated as
\begin{equation}
A_{\lambda} = 1.086 C_{ext} \times l_d
\end{equation}
Integrating the dust column density $\Sigma_{d}$ over the image area \textit{S}  occupied by the dust, 
yields dust mass in solar units as 
\begin{equation}
M_{d} = S \times \Sigma_{d} = S l_d \int^{a+}_{a-} \frac{4}{3} \pi a^{3} \rho_{d} n(a) da
\end{equation}
where $\rho_d$ is the specific grain mass density  $\sim$ 3 g cm$^{-3}$ 
for graphite and silicate grains  (\citealt{Dra84}). The grain size obtained from the
 extinction curve refer to the upper end of the size distribution (\citealt{Goud94b}),  
the upper limit of grain size $a_{+}$ for the program galaxies can be scaled as  
\begin{equation}
a_{+} = \frac{<a>}{a_{Gal}} \times0.22 \mu m 
\end{equation}
where $\frac{<a>}{a_{Gal}}$ is the relative grain size, listed in  Table 5 
and the lower limit $a_{-}$ is taken as 0.005 $\mu$m.
The extinction efficiencies  $Q_{ext}$  for spherical grains composed of graphite and 
silicate with equal abundance   are taken from the published values (\citealt{Jura82, Dra84}).
In the optical region, $Q_{ext}$ can be parametrized as 
\[Q_{ext,silicate} = \left\{
  \begin{array}{l l}
    0.8 a/a_{silicate} & \quad \textrm{for  $a < a_{silicate}$}\\
    0.8& \quad \textrm{for $a \geq a_{silicate}$}
  \end{array} \right.\]
\[Q_{ext,graphite} = \left\{
  \begin{array}{l l}
    2.0 a/a_{graphite} & \quad \textrm{for  $a < a_{graphite}$}\\
    2.0& \quad \textrm{for $a \geq a_{graphite}$}
  \end{array} \right.\]
with $a_{silicate}$ = 0.1 $\mu$m and $a_{graphite}$ = 0.05 $\mu$m. Putting
 these values together, the total dust mass (M$_{d,optical}$) of the 
 program galaxies was calculated and is listed in   
Table 5. While extracting the mean extinction for the sample
galaxies, the  dusty regions with $A_{V}$ $<$ 0.02 were ignored. 
 
\subsection{Dust mass using IRAS flux}

﻿\begin{table}\begin{center}\tiny{
\caption{Derived data from IRAS flux }
\begin{tabular}{|llccl|}
\hline
 Object  & log L$_{B}$&log L$_{IR}$&log M$_{d,IRAS}$&T$_{d}$\\
         &[L$_{\odot}$] &[L$_{\odot}$]& [M$_{\odot}$]&[K]\\ 
\hline

NGC\, 383  & 10.88  & 9.82$\pm$0.23    & 5.19$\pm$ 0.24         & 33 $\pm$ 4  \\
NGC\, 708  & 10.62  & 9.56$\pm$0.26    & 5.07$\pm$ 0.26         & 32$\pm$ 4   \\
NGC\, 720  & 10.52  & NA&& \\   
NGC\, 1052 & 10.27  & 8.97$\pm$0.13    & 4.98$\pm$ 0.08         & 40$\pm$ 1 \\
NGC\, 1167 & 10.89  & 9.62$\pm$0.23    & 6.29$\pm$ 0.34         & 21$\pm$ 2 \\
NGC\, 1199 & 10.48  & NA&&  \\   
NGC\, 1395 & 10.70  & 8.11$\pm$0.27    & 5.49$\pm$ 0.58         & 23$\pm$ 5  \\
UGC\, 2783 & 10.87  & NA&&         \\   
NGC\, 1407 & 10.79  & 8.44$\pm$0.22    & 5.02$\pm$ 0.23         & 31$\pm$ 3   \\
NGC\, 2534 & 10.27  & 9.51$\pm$0.23    & 4.95$\pm$ 0.22         & 35$\pm$ 4   \\
NGC\, 2644 & 9.91   & NA&& \\   
NGC\, 2768 & 10.70  & 8.89$\pm$0.23    & 5.46$\pm$ 0.21         & 31 $\pm$ 3  \\
NGC\, 2851 & 10.33  & NA&& \\   
NGC\, 2855 & 10.14  & 9.2$\pm$ 0.13    & 5.9$\pm$ 0.08         & 27 $\pm$ 1 \\
NGC\, 3065 & 10.62  & 9.59$\pm$0.13    & 4.94$\pm$ 0.07         & 45$\pm$ 1   \\
NGC\, 3115 & 10.17  & NA&&  \\   
NGC\, 3377 & 9.92   & 7.69$\pm$0.25    & 4.53$\pm$ 0.29         & 35$\pm$ 5   \\
M  \, 105  & 10.46  & NA&& \\   
NGC\, 3489 & 9.92   & NA&&  \\   
NGC\, 3607 & 10.21  & NA&&   \\   
NGC\, 3801 & 10.42  & 9.68$\pm$0.14    & 7.53$\pm$ 0.48         & 16$\pm$ 2   \\
NGC\, 4125 & 10.74  & 9.08$\pm$0.14    & 5.25$\pm$ 0.09         & 36$\pm$ 1   \\
NGC\, 4233 & 10.12  & 8.83$\pm$0.22    & 4.67$\pm$ 0.21         & 35$\pm$ 4   \\
NGC\, 4278 & 9.99   & 8.42$\pm$0.09    & 5.43$\pm$ 0.09         & 32$\pm$ 1   \\
NGC\, 4365 & 10.59  & NA&&  \\   
NGC\, 4494 & 10.62  & NA&&  \\   
NGC\, 4552 & 10.40  & 7.3 $\pm$0.19    & 4.98$\pm$ 0.26         & 31$\pm$ 4   \\
NGC\, 4648 & 9.86   & NA&&  \\   
NGC\, 4649 & 10.82  & 8.76$\pm$0.1    & 4.73$\pm$ 0.07         & 44$\pm$ 1 \\
NGC\, 4697 & 10.33  & 7.09$\pm$0.93    & 5.17$\pm$ 0.47         & 34$\pm$ 1   \\
NGC\, 4874 & 11.20  & NA&&  \\   
NGC\, 5322 & 10.78  & 9.05$\pm$0.15    & 5.03$\pm$ 0.11         & 36$\pm$ 2   \\
NGC\, 5525 & 10.58  & NA&& \\   
NGC\, 5812 & 10.40  & NA&&  \\   
NGC\, 5846 & 10.73  & NA&&  \\   
NGC\, 5866 & 10.24  & 9.55$\pm$0.14    & 6.58$\pm$ 0.07         & 30$\pm$ 1 \\
NGC\, 6166 & 11.40  & 10.03$\pm$0.21   & 5.81$\pm$ 0.28         & 24$\pm$ 2   \\
NGC\, 7052 & 10.57  & 9.88$\pm$0.15    & 5.37$\pm$ 0.12         & 32$\pm$ 2  \\
NGC\, 7454 & 10.18  & NA&&  \\   
\hline
\end{tabular}
\label{IR_result}
}
\end{center}
\end{table} 
 Mass of dust within the galaxy can 
also be estimated from the IRAS flux densities by assuming thermal equilibrium. 
The relation between the dust mass and the observed far-IR flux density $I_{\nu}$, is 
\begin{equation} 
 M_{d} = \frac{4}{3} a \rho_d D^{2} \frac{I_{\nu}}{Q_{\nu}B_{\nu}(T_{d)}}
\end{equation}
where $a$ is  grain radius, $\rho_{d}$ is  specific grain mass density,
 $D$ is  distance of the galaxy in Mpc, $Q_{\nu}$ and $B_{\nu}(T_{d})$ are the grain
emissivity and  Planck function at  temperature $T_{d}$ and  frequency
$\nu$, respectively (\citealt{Hil83}). The temperature of dust was calculated 
following  \cite{You86, Pat07}
\begin{equation}
T_{d} = 49(S_{60}/S_{100})^{0.4} 
\end{equation}
where $S_{60}$ and $S_{100}$ are the  flux densities at 60$\mu$m and 100$\mu$m. The calculated dust 
temperature  is listed  in Table 6.  The estimated dust  temperatures should be 
regarded as `representative'  values, since a $range$ of temperature is appropriate for dust 
within early-type  galaxies (\citealt{Goud95}).  The value of T$_{d}$  varies  from 26 K to 39 K 
indicating   possible  presence of `warm' dust in these galaxies. 

 It is shown by \cite{Mat81} that the  observed extinction curve of the Milky Way  
is not consistent with attenuation  of starlight by uniform grain size  and hence the grain 
size $a$ and grain emissivity  $Q_{\nu}$  need  to be  suitably chosen averages. For obtaining 
the average grain  emissivity,  the value for each grain is weighted by the contribution 
of the grain to the flux density and hence by grain volume. Similarly, for a given size 
distribution, an average value of $a$ can also be obtained by weighted average.  
For  particle size distribution function of \cite{Mat77},  a weighted average radius $a$ of 0.1 $\mu$m   was 
calculated by \cite{Hil83}.  
Hence, in  this study  we have taken  average grain size of 0.1 $\mu$m,    
 $\rho_{d}$ = 3 g \ cm$^{-3}$ and $\frac{4a\rho_{d}}{3Q_{\nu}}$ = 0.04 g \ cm$^{-2}$ 
at 100 $\mu$m (\citealt{Hil83}).
The dust mass for galaxies with available IRAS fluxes obtained 
following the method outlined in \cite{Thro86} and \cite{Goud94b} is 
listed in Table 6.

\subsection{Narrow-band images}
A significant fraction ($\sim 75\%$) of our sample was observed with narrowband  H$\alpha$ 
filter for studying properties of  ionized gas. 
 The images obtained through H$\alpha$ filter  include photons
 from H$\alpha$+[NII] lines and from the underlying continuum.
Pure H$\alpha$ line emission map of sample galaxies can be generated as described below.  
 
The  H$\alpha$ and continuum images were first geometrically aligned and 
combined to get the final narrow-band  and  continuum images.
 The image with  better seeing was convolved to match the  psf of narrow-band
and continuum images. A scaled version of the continuum image is subtracted 
from the narrow-band image to get pure H$\alpha$+[NII] emission-line image.
The intensity scaling factor(ISF) was obtained following  the  method 
described by \cite{Spec12}. Under the assumption that mix of foreground 
stars represents the stellar population of program galaxy, mean of 
ratio of   fluxes of the foreground stars  in 
line and continuum images was used as the scaling factor. 
The continuum subtracted images  shows presence of line emission
in many of our sample galaxies.  \cite{Spec12} also cautioned that regardless of 
number of stars taken for estimating ISF, it is very difficult to 
match the stellar population of the observed galaxy with that of the 
foreground stars.

An alternate method to estimate the ISF was also tried. In this method  ellipses 
were fitted to the isophotes of galaxies in both the bands.   
A least  square fit was applied to the intensities
measured in  outer regions of narrow-band and continuum images, which was 
 minimally affected  by the ionized gas.  The slope of this fit gives the scale 
factor between  continuum and emission line images (\citealt{Macch96}).  
Intercept of the least square fit is related to the residual in the sky background 
between the two bands, an accurate sky subtraction reduces the intercept of the 
fit to zero. The accuracy of ISF was checked by examining  the continuum 
subtracted images which are expected to show zero value in the outer regions and 
clean removal of the  foreground stars.  The ISF so obtained were compared with 
those estimated using the foreground stars. For  a good number of galaxies, the two
ISFs agreed well,  in case of discrepant values,  the second method 
being more accurate was preferred.

The resultant images were examined visually for possible presence of  line emission.
Line emission is detected in 23  out of  30 galaxies imaged through narrow-band filter. 
 In most of the line emitting galaxies the 
emission is centered around   the nucleus. The total H$\alpha$+[NII] counts 
N$_{H\alpha+[NII]}$ inside a circular aperture centered on the 
galaxy nucleus
 was  estimated using `DAOPHOT' in IRAF. The aperture size was taken big enough to 
encompass all the line emission from the galaxy. The final aperture sizes
used for the count measurement are given in Table 7. \\
\begin{figure}
\begin{center}  
\includegraphics[totalheight=6.8in,width=5in]{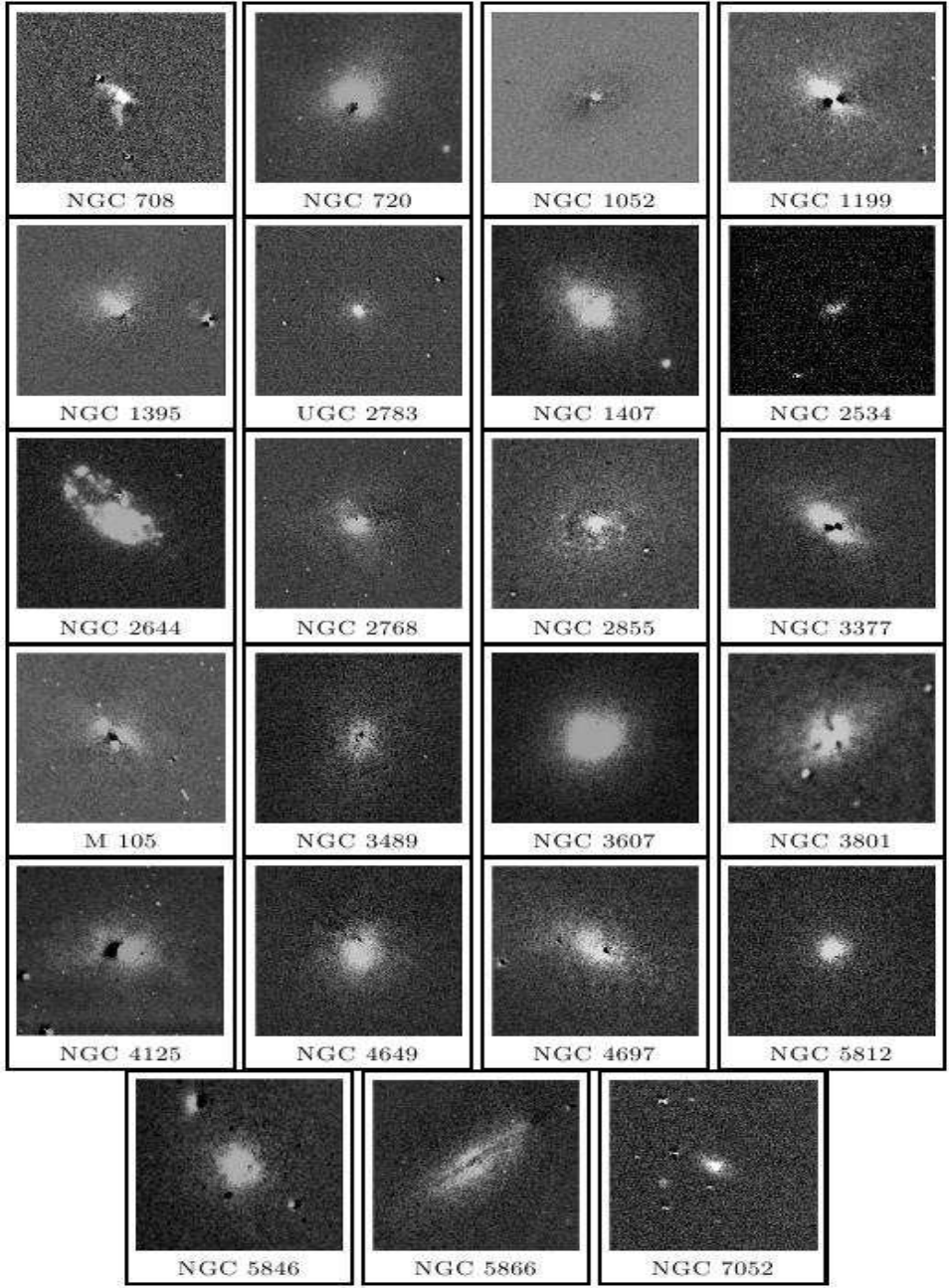}
\caption{${H\alpha}$ Emission Map for sample Galaxies.}
\label{h_emis}
\end{center}
\end{figure}

The line emission maps of our sample galaxies with detectable  line emission 
are  displayed in Figure 4. 
The measured  H$\alpha$+[NII] counts were converted  to  absolute flux scale using the 
following relation
\begin{equation}
 {F_{\alpha}} (erg \ cm^{-2} \ s^{-1}) = C_{\alpha} \times f_{\alpha} (count \ s^{-1})
\end{equation}
where $ C_{\alpha}$ the conversion factor was
determined by comparing the observed counts ($f_{\alpha}$) of the spectrophotometric 
standard star with the expected flux ($F_{\alpha}$) within the bandpass of the
filter used. 
Pure H$\alpha$ flux can be estimated from the observed flux by removing the contamination 
due to [NII] lines using the correction factor 
\begin{equation} 
 Corr. = \frac{1}{1+[NII] Cont.  \times  \frac{[NII]}{H\alpha} \times 1.33}
\end{equation}
where [NII] Cont. is contamination due to the [NII] lines and estimated as the ratio 
of the average transmission efficiency of the filter at $\lambda\lambda$6548, 6583 to 
that of H$\alpha$. The $\frac{[NII]}{H\alpha}$ ratio  for a galaxy is usually determined 
spectroscopically.  In absence of  
spectroscopic information  the value of $\frac{[NII]}{H\alpha}$ was taken as 1.38 (\citealt{Phi86}). 
The derived [NII] contamination,  correction factor and  pure H$\alpha$ flux 
are given  in Table 7. 
The H$\alpha$ luminosities  for the sample galaxies were calculated
using the distance derived from the velocities corrected for local group 
infall onto  Virgo (source Hyperleda) and  Hubble constant 
H$_{0}$ = 73 km sec$^{-1}$ Mpc$^{-1}$ (\citealt{Sper07}). 
  
\begin{table}\tiny{
\begin{center}
\caption{Results for $H\alpha+[NII]$ emission }
\begin{tabular}{|lllllllll|}
\hline
Object&Dist.&Aper.&[NII] &Corr.&$\textit{f}_{H\alpha}$&$L_{H\alpha}$&log $M_{HII}$&Remarks on dust and IG\\
name& Mpc&arcsec& contam- &&$10^{-14}$&$ergs^{-1}$&${M\odot}$&morphology\\\
&&&ination&&$ergcm^{2}s^{-1}$&&\\
\hline   
NGC0708&68.92   & 4  & 0.99 &0.355&0.39&2.22 $\times 10^{41}$&	5.71&	Marginal dust in central region\\	 
NGC0720&22.16   & 29 & 0.79 &0.408&15.7&9.22 $\times 10^{39}$&	4.33&	Nuclear dust \& IG in center\\	  
NGC1052&19.10   & 26 & 0.82 &0.399&22.4&9.77 $\times 10^{39}$&	4.35&	Nuclear dust \& IG in center\\	  
NGC1167&70.23   &    &No Emission&&&&                            &   Dust patch \\	  
NGC1199&35.20   & 32 & 1.32 &0.291&3.5 &5.19 $\times 10^{39}$&	4.08&	Marginal dust in central region\\
NGC1395&21.07   & 20 & 0.80 &0.405&3.91&2.08 $\times 10^{39}$&	3.68&	Small dust lane along minor axis\\
UGC2783&87.39   & 19 & 0.77 &0.414&2.72&2.48 $\times 10^{40}$&	4.76&	IG in center region\\		  
NGC1407&22.57   & 43 & 0.78 &0.411&18.0&1.10 $\times 10^{40}$&	4.40&	IG in center\\	                  
NGC2534&51.74   & 6  & 0.62 &0.468&1.13&3.62 $\times 10^{39}$&	3.92&	Marginal IG in central region\\	  
NGC2592&29.44   &    &No Emission&&&&                            &	DL along major axis \\
NGC2644&26.63   & 41 & 0.94 &0.367&13.3&1.13 $\times 10^{40}$&	4.42&   Dust patch \& IG near center\\
NGC2768&22.52   & 26 & 0.76 &0.418&6.9 &4.19 $\times 10^{39}$&	3.98&   Dust patch \& IG near center\\
NGC2855&24.96   & 42 & 0.94 &0.367&9.7 &7.23 $\times 10^{39}$&	4.22&   Spiral like filaments dust \& IG filaments\\  
NGC3115&8.15    & &No Emission&&&& &   no dust  \\
NGC3377&10.34   & 35 & 0.90 &0.377&7.24&9.26 $\times 10^{38}$&	3.33&   Dust \& IG along major axis\\	     
M105   &13.36   & 26 & 0.90 &0.377&8.03&1.71 $\times 10^{39}$&	3.60&   IG in center\\	                     
NGC3489&10.68   & 25 & 0.73 &0.427&15.2&2.07 $\times 10^{39}$&	3.68&   Warped dust \& IG feature\\		  
NGC3607&14.17   & 35 & 0.90 &0.377&35.0&8.40 $\times 10^{39}$&	4.29&   Dust \& IG ring around center\\	      
NGC3801&49.27   & 35 & 1.1  &0.331&4.64&1.35 $\times 10^{40}$&	4.49&   Multiple DL \& filamentary structure of IG\\ 
NGC4125&22.98   & 49 & 0.76 &0.418&14.1&8.91 $\times 10^{39}$&	4.31&   Small dust lane \& IG in central region\\    
NGC4365&17.84   & 35 & No Emission&&&& &   Dust patch \\ 
NGC4649&16.64   & 32 & 0.80 &0.405&9.79&3.24 $\times 10^{39}$&	3.87&   Nuclear dust; IG in center\\  
NGC4697&2.80    & 49 & 0.81 &0.402&31.8&2.98 $\times 10^{38}$&	2.84&   Dust patch \& IG in center\\		
NGC5525&78.62   &    &No Emission&&&&                            &   Large dust lane along Major axis\\
NGC5812&27.13   & 23 & 0.92 &0.372&25.3&2.23 $\times 10^{40}$&	4.71&   Dust patch \& IG in center\\	      
NGC5846&25.45   & 23 & 0.97 &0.360&18.9&1.46 $\times 10^{40}$&	4.53&   Nuclear dust patch \& IG in center\\     
NGC5866&13.20   & 41 & 0.88 &0.382&10.7&2.23 $\times 10^{39}$&	3.71&   Large dust lane \& associated IG\\	     
NGC6166&132.53  &    &No Emission&&&&                            &   No dust\\		                    
NGC7052&67.53   & 17 & 0.99 &0.355&0.74&4.04 $\times 10^{39}$&	3.97&   Small dust patch \& IG in center\\       
NGC7454&29.04   &    &No Emission&&&&                               &   No dust\\                              
\hline			                                                                    
\end{tabular}
\label{h_result}
\end{center}
}
\end{table}

\subsubsection{Mass of ionized gas}
The mass of  ionized gas M$_{HII}$ was estimated following \cite{Goud94b} 
and \cite{Macch96}, using the relation,  
  
\begin{equation} 
 M_{HII} = 2.33 \times 10^{3} (\frac{L_{H\alpha}}{10^{39}})(\frac{10^{3}}{n_{e}}) M_{\odot}
\end{equation}
where L$_{H\alpha}$ is the H$_\alpha$ luminosity. 
The derived mass of  ionized gas is listed in Table 7. 
The uncertainty in the emission line flux measurement is 
discussed in detail by \cite{Macch96}.  The main source of error is 
the uncertainty involved in  determination of ISF which  depends on the 
equivalent width of the ionized-gas emission. Hence, the measurement error  
is more for the galaxies with faint and diffuse emission.  
However,  other sources like  photon-noise in the
narrow-band and continuum images,  residual spatial variation after flat-fielding,  
error in  sky subtraction, inaccurate psf matching of the narrow-band and continuum 
images etc. also contribute to the final error. Typical error
in our H$\alpha$+[NII] flux measurement is $\sim$ 20\%. 

\section{ Results and Discussion}
\textit{Dust}:   For the galaxies with obvious features of dust extinction, it is
found that  morphology and extent of the dust feature in 
colour-index image is similar to  that of extinction map. This  indicates  that  
the extinction map  approximates the 
 dust obscured region to a high degree. An analysis of  extinction curves for  
12 galaxies is presented  in this work. The 
extinction curves for eight galaxies are   reported for the first time. The value of $R_{V}$ 
and relative grain  size for the sample 
galaxies obtained by analyzing the extinction maps are given in Table 5. 
The value of  $R_{V}$  for  sample galaxies  varies from 2.1 to 3.8 with an average 
of 2.95. It is close to the previously reported values by \cite{Goud94c} ($R_{V}$= 2.7), 
\cite{Pat07} ($R_{V}$= 3.01) and \cite{Fin10d} ($R_{V}$= 2.82). 
We find that the galaxies having  $R_{V}$ values
less than the canonical value of 3.1 have well settled dust morphology in the form of 
 a dust lane or a ring and have relatively smaller  grain  size  than  that of the  Milky Way.
  Conversely, the galaxies with larger values of  $R_{V}$ show irregular 
dust morphology and larger  relative grain size. This result is in good agreement with results of
\cite{Goud94b}, \cite{Pat07} and \cite{Fin10i}. There are four galaxies in common with the samples 
of \cite{Goud94c} and \cite{Pat07}. Table 8  presents the  value of $R_{V}$,  
relative grain size and optical dust mass for the four common galaxies  along with the  
 previously reported  values in the literature.   Except for 
galaxy   \astrobj{NGC 4125}, which shows complex dust structure (\citealt{Goud94c}),  our results are 
in good agreement with those reported in the literature.

\cite{Goud94b}  explored  various dust destruction mechanisms such as  sputtering of 
dust by supernova blast wave, turbulent shocks, thermal ions and  hot gas etc. It is  shown that 
for volatile grain of radius $\sim$ 0.1 $\mu$m, the  lifetime of  dust grain varies from 10$^{7}$ to 
10$^{9}$ years. The mass of dust and its grain size are expected to drop gradually with time 
since the dust was acquired by the galaxy.  In this scenario, the smooth and  regular dust 
lane morphology  associated with smaller grain size is expected, if the galaxy had sufficient time 
for the dust to settle down in a regular shape.  On the other hand if the dust is acquired 
recently, it will lack regular dust  morphology as  in the case of \astrobj{NGC 3801}. 
Among our sample galaxies \astrobj{NGC 3801} shows considerably 
larger relative grain size (1.47) and 
larger value of $R_{V}$. A multiwavelength study of this galaxy by \cite{Hota12} revealed
that this galaxy has   a kinematically decoupled core or an extremely warped gas disc, dust 
filaments with complex structures, 
recent star burst with age less than 500 Myr and  evidence of ionized gas (as seen 
in this study too). All these facts led \cite{Hota12} to suggest that this galaxy
is a merger-remnant early-type galaxy.  The results of \cite{Hota12}
support that  \astrobj{NGC 3801}  acquired dust in the recent past and did not have enough time 
to settle it into a smooth,  regular morphology, which may result in the observed  larger grain
 size and $R_{V}$ value, as suggested by \cite{Goud94c}.  

  The  dust mass reported in this work using optical extinction is in the 
range $10^2 - 10^4$ $M_{\odot}$  while, the dust mass estimated using  IRAS fluxes is   
higher by a factor of $\sim$ 10$^2$. This indicates  that a significant fraction of 
dust is diffusely distributed throughout the galaxy (\citealt{Goud95, Wise96}) which 
could not be detected with optical observations, 
this is in agreement with that  of previous studies (\citealt{Pat07, Fin08, Fin10d}).

\begin{table*}\tiny{
\caption{Comparison of dust properties}
\begin{tabular}{|l|ccl|cccl|}
\hline
Galaxy Name&\multicolumn{3}{|c|}{This Paper}& \multicolumn{4}{|c|}{Reported values for comparison}\\
\hline
& $R_{V}$&$\frac{<a>}{a_{Gal}}$ & log($\frac{M_{d,opt}}{M_{\odot}}$) & $R_{V}$&$\frac{<a>}{a_{Gal}}$&log($\frac{M_{d,opt}}{M_{\odot}}$) & Reference\\
\hline
NGC\,2534&2.1$\pm$0.20&0.83&3.32$\pm$0.03&2.03$\pm$0.28&0.80&3.78&  \cite{Pat07} \\      
NGC\,3489&3.1$\pm$0.33&1.04&2.40$\pm$0.05&3.38$\pm$0.21&1.09&3.66& \cite{Pat07} \\      
NGC\,4125&3.6$\pm$0.27&1.36&2.10$\pm$0.02&2.74$\pm$0.33&0.96&5.22& \cite{Goud94c}\\      
NGC\,5525&3.2$\pm$0.28&1.01&4.04$\pm$0.09&3.15$\pm$0.17&0.99&5.67& \cite{Pat07} \\      
\hline
\end{tabular}
\label{dust_compare}
}
\end{table*}

\textit{Ionized gas}:
H$\alpha$ line emission is detected in $\sim$ 85\% of our sample  galaxies observed in narrow-band, 
  the results are listed in Table 7. 
The H$\alpha$ luminosity of the sample galaxies  
 lies in the range 10$^{38}$ - 10$^{41}$ {\it erg sec$^{-1}$} and the mass of ionized gas
lies in the range $\sim$  10$^{3}$ - 10$^{5}$ M$_{\odot}$.  The ionized gas and dust have identical
morphology  in almost all the  galaxies detected in H$\alpha$ (refer Figure 2 and 4). 
We have several galaxies in common with earlier studies by  \cite{Kim89}, \cite{Trin91}, 
\cite{Shi91}, \cite{Goud94b}, \cite{Macch96} and  \cite{Fin10i}.
The H$\alpha$ flux estimated in this work are in good agreement with the flux 
reported in the literature (refer Table 9).
There are 6 galaxies for 
which H$\alpha$ flux are available from
more than one source and a significant scatter is noticed among them. 
The source of scatter may be due to the  higher uncertainty involved in the method of 
determination of the line emission fluxes as discussed in Section 3.3. Similar 
discrepancy in  H$\alpha$ flux was also reported by \cite{Fin10i}. 

\begin{table}\tiny{
\begin{center}
\caption{Comparison of ${H\alpha}$ fluxes}
\begin{tabular}{|llll|}
\hline
 Galaxy &$\textit{f}_{H\alpha}$& Other obs. & Reference \\
 name& This paper &&\\
&$ergcm^{2}s^{-1}$&$ergcm^{2}s^{-1}$&\\
\hline     
NGC0708& 39$\times 10^{-14}$  &49$\times 10^{-14}$& \cite{Fin10i}\\
NGC0720& 15.7$\times 10^{-14}$&$<$ 1.49$\times 10^{-14}$& \cite{Goud94b}\\
&&$< 3.53 \times 10^{-14}$& \cite{Shi91}\\
&&17.66$\times 10^{-14}$& \cite{Macch96}\\
NGC1052& 22.4$\times 10^{-14}$&8.1$\times 10^{-14}$& \cite{Kim89}\\
NGC1199& 3.5$\times 10^{-14}$ &4.9$\times 10^{-14}$& \cite{Fin10i}\\
NGC1395& 3.91$\times 10^{-14}$&2.2$\times 10^{-14}$& \cite{Goud94b}\\
&&22.7$\times 10^{-14}$& \cite{Macch96}\\
&&9.33$\times 10^{-14}$& \cite{Trin91}\\
NGC1407& 18.0$\times 10^{-14}$&0.9$\times 10^{-14}$& \cite{Goud94b}\\
&&$<$5.11$\times 10^{-14}$& \cite{Shi91}\\
&&4.82$\times 10^{-14}$& \cite{Macch96}\\
NGC2534& 1.13$\times 10^{-14}$&6.4$\times 10^{-14}$& \cite{Fin10i} \\
NGC3377& 7.24$\times 10^{-14}$&11.9$\times 10^{-14}$& \cite{Goud94b}\\
NGC3489& 15.2$\times 10^{-14}$&51.89$\times 10^{-14}$& \cite{Macch96}\\
NGC3607& 35.0$\times 10^{-14}$&7.83$\times 10^{-14}$& \cite{Macch96}\\
NGC4125& 14.1$\times 10^{-14}$&39.5$\times 10^{-14}$& \cite{Goud94b}\\
&&16$\times 10^{-14}$& \cite{Kim89}\\
NGC4649& 9.79$\times 10^{-14}$&11$\times 10^{-14}$& \cite{Trin91}\\
NGC4697& 31.8$\times 10^{-14}$&29.5$\times 10^{-14}$& \cite{Goud94b}\\
&&4.7$\times 10^{-14}$& \cite{Trin91}\\
NGC5812& 25.3$\times 10^{-14}$&5.83$\times 10^{-14}$&\cite{Macch96}\\
NGC5846& 18.9$\times 10^{-14}$&16$\times 10^{-14}$& \cite{Trin91}\\
&&28.02$\times 10^{-14}$& \cite{Macch96} \\
NGC7052& 0.74$\times 10^{-14}$&6.0$\times 10^{-14}$& \cite{Fin10i}\\
\hline       
\end{tabular}
\label{h_compare}\end{center}
}
\end{table}

\subsection{Association of dust and ionized gas} 
Association of dust and ionized gas and their possible origin 
in  E/SO galaxies were  studied recently by \cite{Fin12}.
The HST observations revealed presence of nuclear and
filamentary/patchy dust in the central part of a  significant fraction of early-type galaxies 
(\citealt{Tran01, Mart04}). Most of the galaxies from our sample 
show dust in their central part.  A study similar to that of \cite{Fin12} 
was carried out, by including early-type galaxies  with nuclear, filamentary/patchy dust
to the sample of  galaxies having dust lanes.
 Various relevant parameters like optical dust mass, IRAS dust mass, ionized gas mass, 
optical luminosity, IR luminosity, H$\alpha$ luminosity etc. were collected  from  
\cite{Bus93}, \cite{Goud94b}, \cite{Goud95}, \cite{Sahu96}
 \cite{Sahu98}, \cite{Dew99}, \cite{Sing95}, \cite{Tran01}, 
\cite{Pat07}, \cite{Fin10i}, \cite{Fin12}, this we denote as combined sample. 
Wherever necessary the quantities have been normalized for  H$_0$  = 73 km sec$^{-1}$ Mpc$^{-1}$.

The lower optical depth and smaller sampling area in galaxies with patchy
/filamentary dust makes it difficult to determine dust mass using extinction values.
Dust mass using  optical extinction could be estimated only for 12 galaxies 
with prominent dust features in our sample.  
It was demonstrated by \cite{Tran01}  that even in the case of 
patchy/filamentary dust, the dust mass determined 
using  optical extinction and IRAS fluxes are well correlated. 
Further, the ionized gas mass and  optical dust mass are found to be 
correlated in early-type galaxies (\citealt{Fer99, Fin12}). 
The slope of log(M$_{HII}$) versus  log(M$_{dust, opt}$) 
for the combined sample was found to be $\sim$ 0.6, which is shallower than the slope ($\sim$ 0.8)
found by  \cite{Fin12} for dust lane galaxies.  This indicates that in the prominent dust 
lane galaxies the source of ionization may be more efficient  compared to the galaxies 
with patchy/filamentary dust.   

In Figure 5 dust mass derived using IRAS fluxes is plotted versus ionized gas mass. A
  good correlation between these two quantities is seen. The best fitting 
line to the observed data points with  slope $\sim$ 0.72 is also shown in the same figure.
Our results about the co-existence of dust and ionized gas and their correlation are 
consistent with those  of similar work by \cite{Goud94b}, \cite{Goud95}, 
\cite{Macch96} and  \cite{Fin08, Fin10d}.

\begin{figure}
\begin{center}
\includegraphics[totalheight=3.4in,width=3.4in]{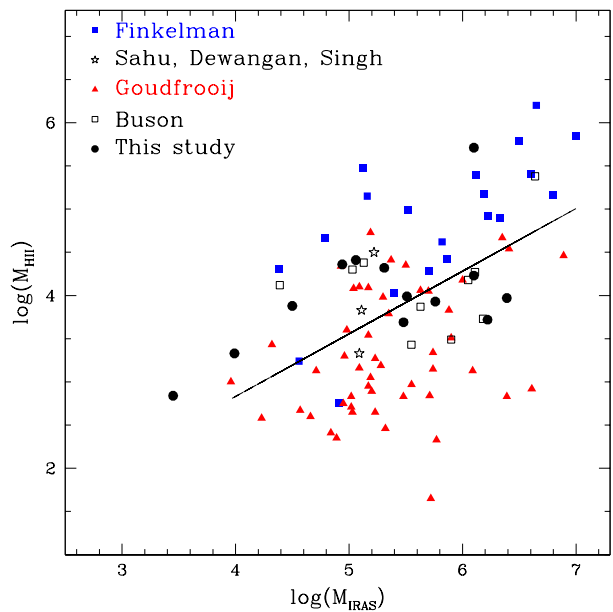}
\caption{ Ionized gas mass  plotted versus  dust mass estimated from $IRAS$ fluxes}
\label{mass_ih}
\end{center}
\end{figure} 

\subsection{Possible source of ionization in early-type galaxies}
The  common presence of ionized gas in early-type galaxies is well established 
(\citealt{Kim89, Goud94b, Macch96, Sarzi06, Fin08}), however, the source of 
ionization is still debated.  Possible sources of ionization in 
early-type galaxies were investigated (\citealt{Mart04, Sarzi10}). 
\cite{Sarzi10}  explored various possible sources of ionization e.g. 
post-asymptotic giant branch  (pAGB) stars, presence of  AGN's, fast 
shocks, OB stars  and interaction with hot ISM.   
On the basis of their ionizing balance arguments the pAGB stars were regarded
 as the best candidate for photoionization; it may be either associated to old 
stellar population or to recent  star formation. However,  on-going star formation
 is also responsible for  ionization of gas in at least  $\sim$ 10\% of their sample. 

The GALEX ultraviolet data revealed that the strong 
UV flux from  $\sim$ 30\%  of nearby bright early-type galaxies can not be explained without 
invoking $\sim$ 1 - 3\% (in total stellar mass) star formation rate in the last billion years.
The fraction of galaxies showing recent star formation could be as high as $\sim$ 50\%, if 
corrected for extinction and proper care is taken for  UV flux contribution of AGNs (\citealt{Kav06}).  

Early-type galaxies are known to have  cold molecular gas (\citealt{Huch95, Mor06, Com07, Oos10}). 
More recently,  in the ATLAS$^{3D}$ sample of early-type galaxies the 
detection rate of molecular gas is $\sim$ 22\%  with H$_{2}$ mass 
$\log$ M(H$_{2}$)/M$\odot$  in the range  7.10 -  9.29. A strong correlation 
between  presence of molecular gas and dust, blue features and young stellar ages seen in the
H$\beta$ absorption indicate 
that the detected molecular gas is often involved in the star formation (\citealt{You11}). 
Based on the CO emission in a representative sample of early-type galaxies, \cite{Com07}
and \cite{Croc11} showed that the  CO-emitting early-type galaxies form a low 
star formation rate (SFR) extension to the empirical law in spirals. 


 We searched for available data on CO emission in our sample galaxies. 
Twelve  out of 26 galaxies with dust and $H{\alpha}$ emission  in our sample 
were mapped  to check for CO emission while for remaining  14 galaxies we don't 
have any information. CO was  detected in only four galaxies namely 
NGC 2768, NGC 3489 (\citealt{Croc11})  NGC 3607, NGC 5866 (\citealt{Davis11})
and upper limits are available for NGC 4125 (\citealt{wiklind95}), NGC 4649 (\citealt{Young02}).
Using [OIII]/H$\beta$ ratio \cite{Croc11} have investigated possible source of
ionization in his sample galaxies. It is concluded that 
in NGC 2768 old pAGB stars/AGN and  in NGC 3489 young pAGB stars are the 
main source of the ionizing photons. However, to get a better estimate of  
 fraction of CO emitting galaxies, where star formation is the dominant 
source of ionization, a detailed investigation similar to that done by \cite{Croc11} 
is required. 

\subsection{Origin of dust and gas in early-type galaxies}
\begin{figure}\begin{center}
\includegraphics[totalheight=3.4in,width=3.4in]{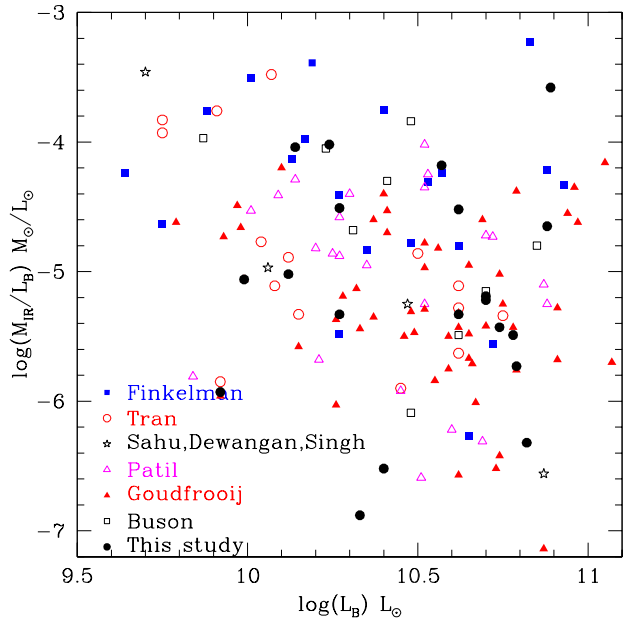}
\caption{Comparison IRAS dust mass, normalized by the luminosity}
\label{mass_lum}
\end{center}
\end{figure} 
In  early-type galaxies presence of ISM in various forms is well established, but its 
origin is still an open issue.
Many evidences support their co-existence and physical association, 
 pointing  towards a common origin (\citealt{Goud94b, Goud95, Macch96, Sarzi06, Fin10i, Fin12}).
 As discussed earlier, $H{\alpha}$ emission is detected in a significant fraction of 
our sample galaxies.
The matching morphology of line emitting regions with that of optical extinction
indicates co-existence of dust and ionized  gas in them.
The possible sources for origin of dust and gas in these systems are internal and external.
 In  case of internal origin dust is deposited through  mass loss from  evolved stars.      
The dust particles  also get destroyed due to  sputtering in various
environments. It  has been demonstrated that the mass of dust observed in early-type galaxies is 
generally higher than that expected from stellar mass loss, indicating  an 
external origin of dust such as galaxy interaction or mergers (\citealt{Goud95, Pat07, Fin12} and references therein).

 To investigate possible  origin of dust in early-type galaxies
relationship between optical luminosity 
and  dust mass normalized by  luminosity is  also explored 
(\citealt{Goud95, Fer99, Pat07}).  In Figure 6, 
  IRAS dust mass normalized with respect to 
blue luminosity  is plotted against  blue luminosity for the combined sample. 
The absence of any apparent correlation between these quantities suggests the 
possibility of external  such as merger related origin of dust and gas in these galaxies. 

The observed misalignment between the angular momentum vectors of
interstellar gas and stellar system in several early-type galaxies 
suggests that they are kinematically
decoupled (\citealt{Ber88, Kim89, van95, Caon00, Kraj08, Sarzi06}). 
Further, \cite{Sarzi06} showed that the kinematical misalignment between the gaseous and 
stellar component is  strongly dependent on  
apparent flattening and the level of rotational support in  these galaxies. The flatter and
faster rotating early-type galaxies are known to have preferentially co-rotating gaseous and 
 stellar systems, indicating that  an internal origin of gas may play an important role in 
fast rotating galaxies. Hence, it is necessary  to invoke a balance between the internal and 
external origin of gas and dust to  explain the recent  observational results (\citealt{Fin12}). 
Moreover, the observed blue  colours  (near-UV - $r$ $<$ 5.5) in a large sample of 
early-type galaxies can only be explained by introducing some amount of recent star formation
in these galaxies (\citealt{Kav06}).
The  recycled gas from stellar mass loss within the galaxy is not enough  
to account for their observed   blue colour.  Hence,  an additional source of  fuel 
 for star formation is unavoidable, may be from  external origin. 

\section{Summary}
We have explored  the properties of dust and ionized gas in a sample of 40 nearby
early-type galaxies using broad-band and narrow-band imaging data. The main findings of 
this work are summarized below \\
\begin{enumerate}
 \item The value of $R_V$, the ratio between the total extinction in $V$ band and
the selective extinction $E(B-V)$ between $B$ and $V$ bands, for the
sample galaxies derived using our observations lies  in the range 2.1 - 3.8 
with an average of 2.95. 
\item  The extinction curves derived for the sample galaxies run parallel to
that of the canonical curve of our Galaxy, suggesting similar optical extinction
properties of dust grains in extragalactic environment as that of the Milky Way.
The relative size of the dust grains $<a>/a_{Gal}$ in these galaxies is found to vary
between 0.75 - 1.47. 
\item  Mass of dust calculated using the total optical extinction is in the
range 10$^{2}$ to 10$^{4}$ M$\odot$,  lower than those estimated using
IRAS fluxes by a factor of $\sim$ 10$^{2}$. This indicates that a  significant
fraction of dust is diffusely distributed,
throughout the galaxy and remains undetected in optical observations. 
\item  The mass of the ionized gas lies in the range  7x10$^{2}$  - 5x10$^{5}$ M$\odot$.
The morphology of dust and ionized gas emission for these galaxies is
found to be similar, implying their plausible physical association. 
Among various mechanism, at least in some of the early type galaxies star formation at a low level could
be a major source of ionization, however more CO observations of H$\alpha$ emitting early-type galaxies
are necessary to determine the source of the ionization in these galaxies.
\item The absence of any apparent correlation between the blue luminosity and 
normalized IRAS dust mass  suggests the probability of   
merger related origin of dust and gas in these galaxies, over and above the  internal origin 
through stellar mass loss.
\end{enumerate}
\section*{Acknowledgments}
We thank the anonymous referee for valuable comments which improved the 
scientific contents of the paper.
SK, LKC and SKP are grateful to ISRO for financial support  
under RESPOND scheme (project no. ISRO/RES/2/343/2007-08). 
We thank the directors of IIA, Bangalore, IUCAA, Pune and the Time Allocation
Committees for allotting the nights for our observations. SK is greatful to Prof. T.P. Prabhu and Prof. A.K. Kembhavi, for the local hospitality and use of
computational facilities at CREST(IIA) and IUCAA, respectively. 
We acknowledge  Dr. M. K. Patil and Dr. Sudhanshu Barway for their 
valuable suggestions. The authors are thankful to Late Prof. R.K. Thakur for his constant 
encouragement and moral support throughout this work. 
 This research has made use of the NASA/IPAC Extragalactic Database (NED) which is 
operated by the Jet Propulsion Laboratory, California Institute of Technology, under 
contract with the National Aeronautics and Space Administration. 
This research has made use of NASA's Astrophysics Data System. We acknowledge 
the usage of the HyperLeda database (http://leda.univ-lyon1.fr).

\end{document}